\documentclass[letterpaper,10pt,nofootinbib,aps,tightenlines,twocolumn]{revtex4}

\usepackage{amsmath,amsfonts,amssymb}
\usepackage{mathrsfs}
\usepackage{graphicx}
\usepackage[english]{babel} 
\usepackage{color}
\usepackage{array}
\usepackage{appendix}

\usepackage{epstopdf}

\def\hatn{\mathbf{\hat n}}

\newcommand{\beq}{\begin{equation}}
\newcommand{\eeq}{\end{equation}}
\newcommand{\bga}{\begin{gathered}}
\newcommand{\ega}{\end{gathered}}
\newcommand{\beqa}{\begin{eqnarray}}
\newcommand{\eeqa}{\end{eqnarray}}

\begin{document}
\title{Patchy Screening of the Cosmic Microwave Background by Inhomogeneous Reionization}
\author{Vera Gluscevic$^1$, Marc
     Kamionkowski$^{2}$, and Duncan Hanson$^{3}$}
\affiliation{$^1$California Institute of Technology, Mail Code 350-17,
     Pasadena, CA 91125, USA\\
     $^2$Department of Physics and Astronomy, Johns
     Hopkins University, Baltimore, MD 21218, USA\\
     $^3$Department of Physics, McGill University,
     Montreal QC H3A 2T8, Canada} 
\date{\today}
\begin{abstract}
We derive a constraint on patchy
screening of the cosmic microwave background from
inhomogeneous reionization, using off-diagonal
$TB$ and $TT$ correlations in WMAP-7 temperature/polarization data.
We interpret this as a constraint on the rms optical-depth
fluctuation $\Delta\tau$ as a function of a coherence multipole
$L_C$.  We relate these parameters to a comoving coherence
scale, of bubble size $R_C$, in a phenomenological model where
reionization is instantaneous but occurs on a crinkly surface,
and also to the bubble size in a model of ``Swiss cheese"
reionization where bubbles of fixed size are spread over some
range of redshifts.  The current WMAP data are still too weak, by
several orders of magnitude, to constrain reasonable models, but
forthcoming Planck and future EPIC data should begin to
approach interesting regimes of parameter space.  We also present
constraints on the parameter space imposed by the recent results
from the EDGES experiment. 
\end{abstract}
\maketitle

\textit{Introduction---}Simple estimates have long shown that radiation from the first
star-forming galaxies in cold-dark-matter models should reionize
the intergalactic medium (IGM) at redshifts $z\sim 10$
\cite{Kamionkowski:1993aw}, but the details of this epoch of
reionization (EoR) are still unclear.  Quasar observations suggest
that the tail end of reionization occurred at a redshift
$z\gtrsim6$ \cite{reionquasar}, but the implications of these
measurements are difficult to interpret precisely
\cite{Oh:2004rm}.  A constraint $\tau = 0.074 \pm 0.034$ to the
optical depth to re-scattering of cosmic-microwave-background
(CMB) photons suggests a reionization redshift $z=10.6\pm1.4$
\cite{Komatsu:2010fb,Larson:2010gs} if
reionization occurred everywhere suddenly.  However, the current
paradigm is that reionization was highly inhomogeneous
\cite{Barkana:2000fd}: regions of ionized gas, seeded by the
first light sources, grew and eventually coalesced, filling the
entire volume of the IGM.  If so, then
reionization may have occurred over some redshift range.  Recent
searches for the  redshifted 21-cm signal from neutral hydrogen
\cite{Furlanetto:2006jb,Pritchard:2011xb} now suggest that 
reionization occurred over a redshift range $\Delta z\gtrsim
0.06$ \cite{Bowman:2012hf}.  The South Pole Telescope
collaboration has now bounded that redshift range from above to
be $\Delta z\lesssim 7.9$ \cite{Zahn:2011vp}, by searching for
a fluctuating kinetic-Sunyaev-Zeldovich (kSZ) signal
\cite{Gruzinov:1998un}, under assumptions that the
mechanism of reionization accords with prevailing theoretical
models.  

Here we study the effects of patchy screening on the CMB
\cite{Dvorkin:2009ah}. Thomson scattering
of CMB photons during the EoR damps small-scale CMB fluctuations by
a factor $e^{-\tau(\hatn)}$, where $\tau(\hatn)$ is the
optical depth in direction $\hatn$ on the sky.  Patchy screening
gives rise to a direction-dependent optical depth $\tau(\hatn)$.
This then produces a B-mode polarization that is correlated in a
characteristic way with the temperature and with the E mode
polarization \cite{Dvorkin:2008tf,Dvorkin:2009ah}, and it also
modulates the power in the temperature map.
Here we interpret prior null searches for a modulation of CMB
power \cite{Hanson:2009gu} in terms of an upper limit to
optical-depth fluctuations, and we apply a minimum-variance
estimator \cite{Dvorkin:2009ah} for $\tau$ fluctuations to the
WMAP-7 temperature and
polarization maps \cite{lambda} to search for patchy screening by measuring the off-diagonal $TB$ correlations.
We derive an upper limit to all multipoles of the power spectrum
$C_L^{\tau\tau}$ up to $L=512$.  We then discuss implications of
these constraints for a simple phenomenological reionization
model whose parameters might serve as figures of merit for
future experiments. We revisit predictions for future
experiments and discuss constraints on the parameter space
imposed by the recent results from the EDGES
\cite{Bowman:2012hf} experiment.

\textit{Formalism---}Patchy screening suppresses primary
anisotropies (marked with 
tilde), so the observed temperature fluctuation and polarization
are, respectively,
\beq
\bga
\Delta T(\hatn) = e^{-\tau(\hatn)}\widetilde{\Delta T}(\hatn),\\
p(\hatn) \equiv Q(\hatn)+iU(\hatn) = e^{-\tau(\hatn)}\widetilde{p}(\hatn),
\ega
\label{eqn:tau_suppression}
\eeq
where $Q$ and $U$ are the usual Stokes parameters. 
All temperature and polarization correlations in the CMB can, in
principle, be used to reconstruct the map of $\tau(\hatn)$. The
$EB$ estimator will ultimately provide the best sensitivity to
patchy screening \cite{Dvorkin:2008tf,Gluscevic:2009mm}, once
low-noise polarization measurements are available with future
CMB exeriments. With WMAP and Planck \cite{Planck}, however, the best sensitivity is
achieved with the $TT$ correlation, which we discuss below.  Here,
we derive a constraint to patchy screening from the $TB$ correlation, as a proof of principle.
The estimator for the optical-depth fluctuation is
\cite{Dvorkin:2008tf, Gluscevic:2012me},
\begin{eqnarray}
\widehat\tau_{LM} &=&- i N_L \int d\hatn\, {Y_{LM}(\hatn)} \nonumber\\
 & \times& \left[\sum_{lml'm'} {\bar B_{lm}^*} {}_2Y_{lm}(\hatn)
 \widetilde{C}_{l'}^{TE}\bar T_{l'm'}
{}_2Y_{l'm'}^{*}(\hatn) + \mathrm{cc} \right],
\label{eqn:taulm_estimator}
\end{eqnarray}
where $Y_{lm}$ and ${}_2Y_{lm}$ are spherical harmonics and
spin-weighted spherical harmonics, respectively, and the sum is
only over $l+l'+L=$ odd. The unbarred $B_{lm}$ and $T_{lm}$ are the
observed temperature and polarization multipoles, recovered from
the maps and corrected for the combined instrumental-beam and
pixelization transfer function $W_l$; bars represent the
inverse-variance--filtered (IVF) multipoles, $\bar B_{lm} \equiv
B_{lm}/C_l^{BB}$ and $\bar T_{lm} \equiv T_{lm}/C_{l}^{TT}$,
where the $TT$ and $BB$ power spectra are analytic estimates of
the total (signal plus noise) power spectrum  $C^{XX}_l \equiv
\widetilde C^{XX}_l + C_l^{XX,\text{ noise}}W^{-2}_l$ in a given
frequency band,
for $XX\in \{TT,BB\}$. The normalization $N_L$ can be
calculated either analytically or using Monte Carlo
simulations. The estimator is equivalent to the real part of the
cosmic-birefringence estimator in
Refs.~\cite{Gluscevic:2009mm,Gluscevic:2012me},
the only difference being the parity
condition.  Ref.~\cite{Gluscevic:2012me} demonstrated that
the full-sky formalism with the full-sky IVF procedure
described above is justified in spite of the sky cuts introduced
by masking the Galaxy.

\begin{figure}[htbp]
\includegraphics[height=7cm,keepaspectratio=true]{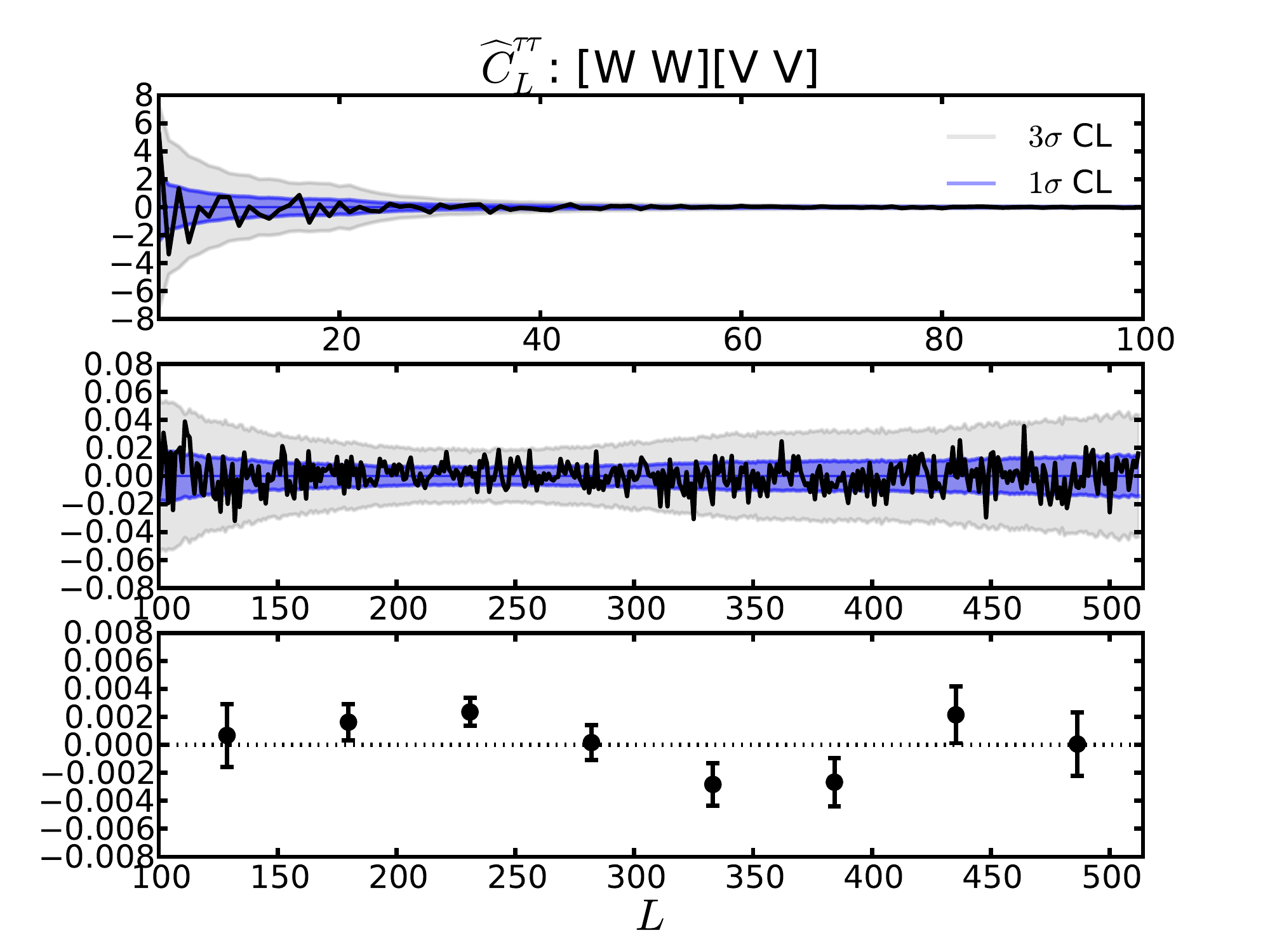}
\caption{The measurement of the power spectrum of fluctuations
     of the optical depth $\tau$ from $TB$ correlation with corresponding
     $1\sigma$ and $3\sigma$ confidence-level intervals for all multipoles up to the resolution limit of WMAP-7 is shown in the top two panels. A binned
     version with associated statistical uncertainty is shown in the bottom panel. The
     first two bins are $-0.0085\pm 0.1264$ at $L=26$, and
     $0.0029\pm 0.0056$ at $L=77$; they are omitted for the
     sake of clearer presentation. The measurements are
     consistent with zero at all multipoles.\label{fig:ctautau}}
\end{figure}
The $TB$ correlations sought by this estimator can in principle also be generated by
re-scattering of CMB photons and by the kSZ effect from re-scattering. However,
Ref.~\cite{Dvorkin:2008tf} showed that the estimator is
relatively insensitive to the kSZ effect, and also that only the
large-scale ($l\lesssim$40) temperature fluctuations are
sensitive to the former mechanism. In order to avoid large-scale contamination from pixel-pixel
noise correlations in WMAP, we discard $T_{lm}$ and $E_{lm}$
multipoles below $l=100$ from our analysis anyway, so we
effectively probe only patchy screening.

The estimator for the corresponding power spectrum of
fluctuations of $\tau$ is
\beq
     C_L^{\widehat\tau \widehat\tau} \equiv
     [f_\text{sky}(2L + 1)]^{-1}\sum\limits_M {{{\widehat
     \tau }_{LM}}\widehat \tau _{LM}^*},
\label{eqn:ctautau_estimator}
\eeq
where $f_\text{sky}$ represents the fraction of the sky admitted
by the analysis mask, correcting for the fact that the full-sky
analysis is applied to the maps where a portion of the pixel
values (mostly around the Galactic plane) was set to zero.
When evaluated for the fixed cosmology of
Ref.~\cite{Komatsu:2010fb} and for the noise levels appropriate
for the experiment in consideration, this four-point correlation provides a biased
estimate of $C^{\tau\tau}_L$, where the bias mostly arises from the
inhomogeneous pixel noise and the sky cuts. However, if this
trispectrum is estimated by
cross-correlating the $\widehat\tau_{LM}$ signal estimated from
one frequency band with the same signal estimated from another
frequency band, the largest contribution to its bias 
vanishes, because the instrumental noise is uncorrelated in different frequency
bands. The leftover bias can be evaluated and subtracted by
running a suite of null-hypothesis (no patchy signal) Monte
Carlo simulations.  We also use the null simulations to recover
the statistical uncertainty for each measurement following the
procedure described in Ref.~\cite{Gluscevic:2012me}.

\begin{figure}[htbp]
\includegraphics[height=7cm,keepaspectratio=true]{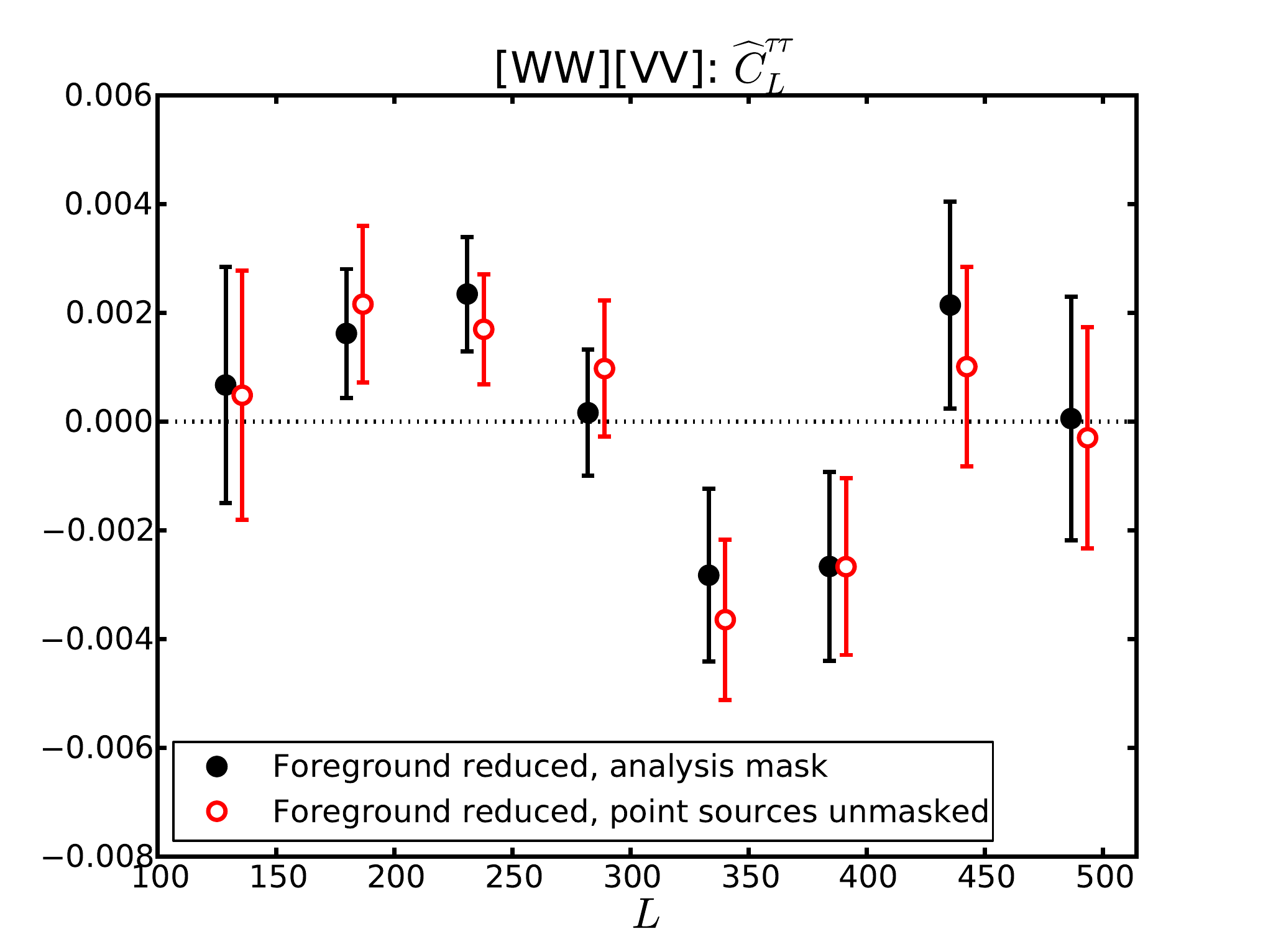}
\caption{Measurement of $\widehat C_L^{\tau\tau}$ from $TB$ correlation in WMAP-7 data. Results shown
     in black (filled circles) are obtained by using the
     analysis mask that covers all the point sources brighter
     than $\sim 1$~Jy, while the results in red (empty circles)
     are obtained after unmasking all the point sources. In
     spite of the large difference in the
     source contamination, the two results differ by much less
     than the statistical uncertainty, and no overall bias is observed.\label{fig:ps_test}}
\end{figure}
\begin{figure}[htbp]
\includegraphics[height=7cm,keepaspectratio=true]{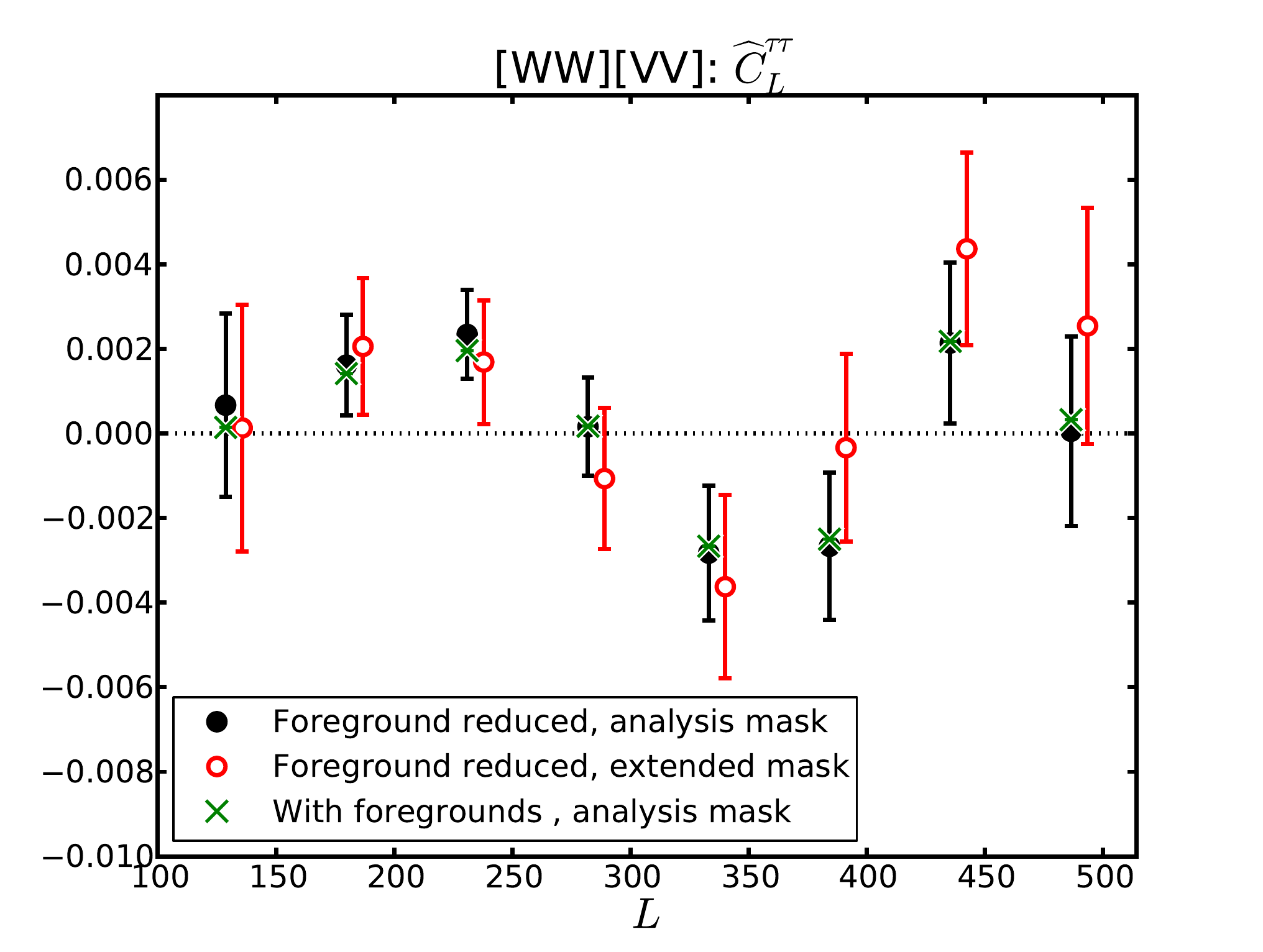}
\caption{Measurement of $\widehat C_L^{\tau\tau}$ from $TB$ correlation in WMAP-7 data. Black filled
     circles represent the measurements obtained from the
     foreground-reduced maps after applying the fiducial
     analysis mask (the fiducial result of Fig.~\ref{fig:ctautau}). The rest of the data points correspond to the two test cases:
     the green x's are obtained from the maps prior to
     foreground subtraction, but using the fiducial mask, while
     the red empty circles are measurements obtained from
     foreground-reduced maps after applying an extended
     mask. No overall bias is observed in the two cases, and all three results are consistent, within the estimated statistical uncertainty.\label{fig:fg_test}}
\end{figure}

\textit{Results from TB estimator---}We only show results for the cleanest
band-cross-correlation [WW][VV] where the estimate of
$\tau_{LM}$ recovered from the W band is cross-correlated with
the estimate from the V band. 
Prior to the
analysis, we mask out the Galaxy and the known point sources
using the fiducial 7-year analysis masks available at the LAMBDA
website \cite{lambda} (where for the combined mask
$f_\text{sky}\simeq68\%$). After subtracting the bias, we
recover a de-biased estimate $\widehat C_{L}^{\tau\tau}$ of the
power spectrum at each multipole up to $L=512$; Fig.~\ref{fig:ctautau} shows the
binned measurements with estimated uncertainties.
At all multipoles, we recover consistency with zero within
the $3\sigma$ confidence level.  

\textit{Tests of Systematics---}Our simulations do not include
polarized point sources nor foreground residuals. In order
to test their impact on our estimates of the power spectrum and
associated statistical uncertainty, we perform the tests described in
Ref.~\cite{Gluscevic:2012me}. The results of these tests are
shown in Figs.~\ref{fig:ps_test} and \ref{fig:fg_test}, which demonstrate that the foregrounds and point sources do not significantly affect the results reported in
Fig.~\ref{fig:ctautau}.  

\begin{figure}[htbp]
\includegraphics[height=7cm,keepaspectratio=true]{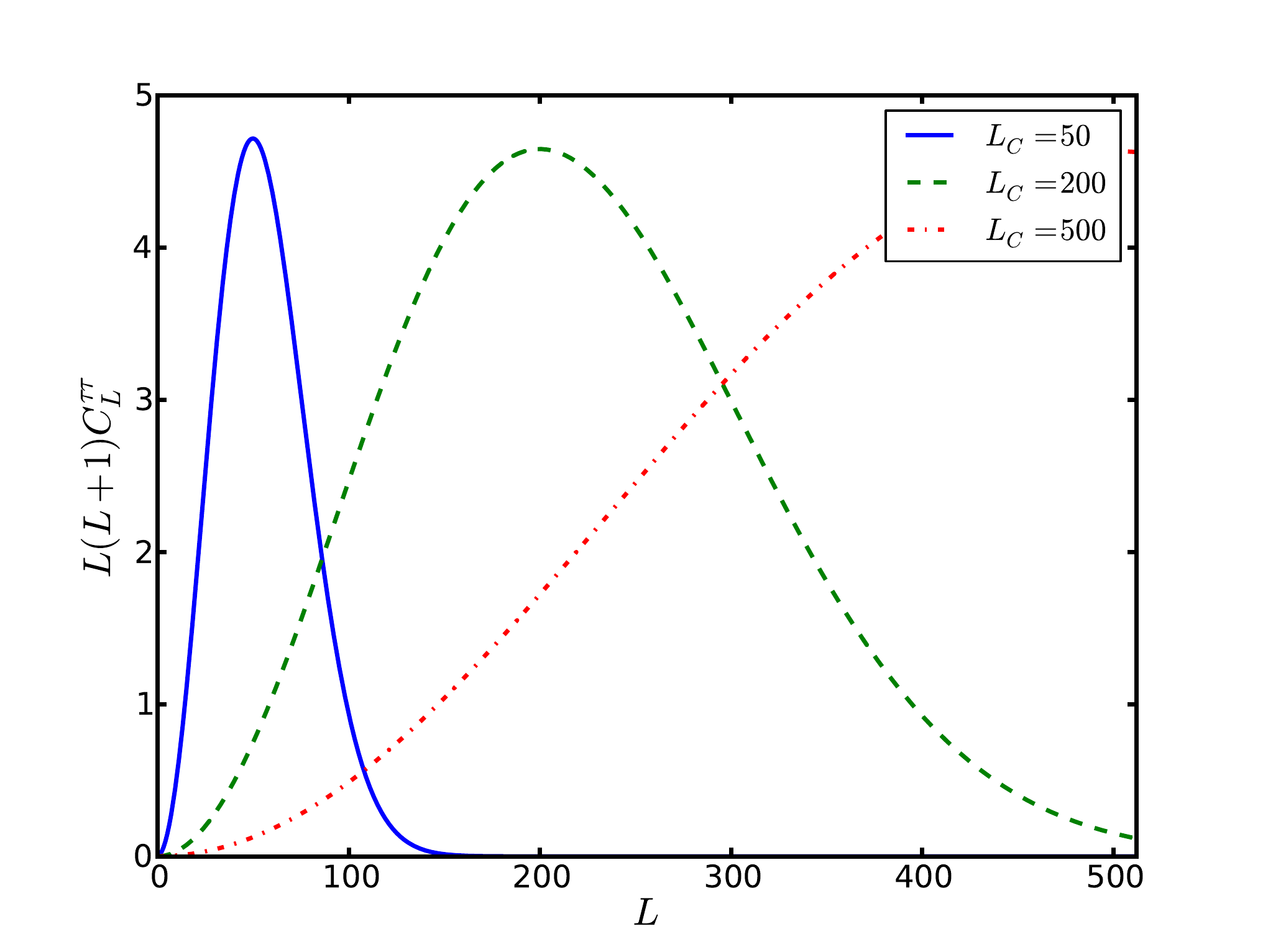}
\caption{Family of simple patchy-reionization models, given by
     Eq.~\eqref{eq:models}, for $\Delta\tau=1$, and different values of $L_C$.\label{fig:models}} 
\end{figure}
\begin{figure}[htbp]
\includegraphics[height=7cm,keepaspectratio=true]{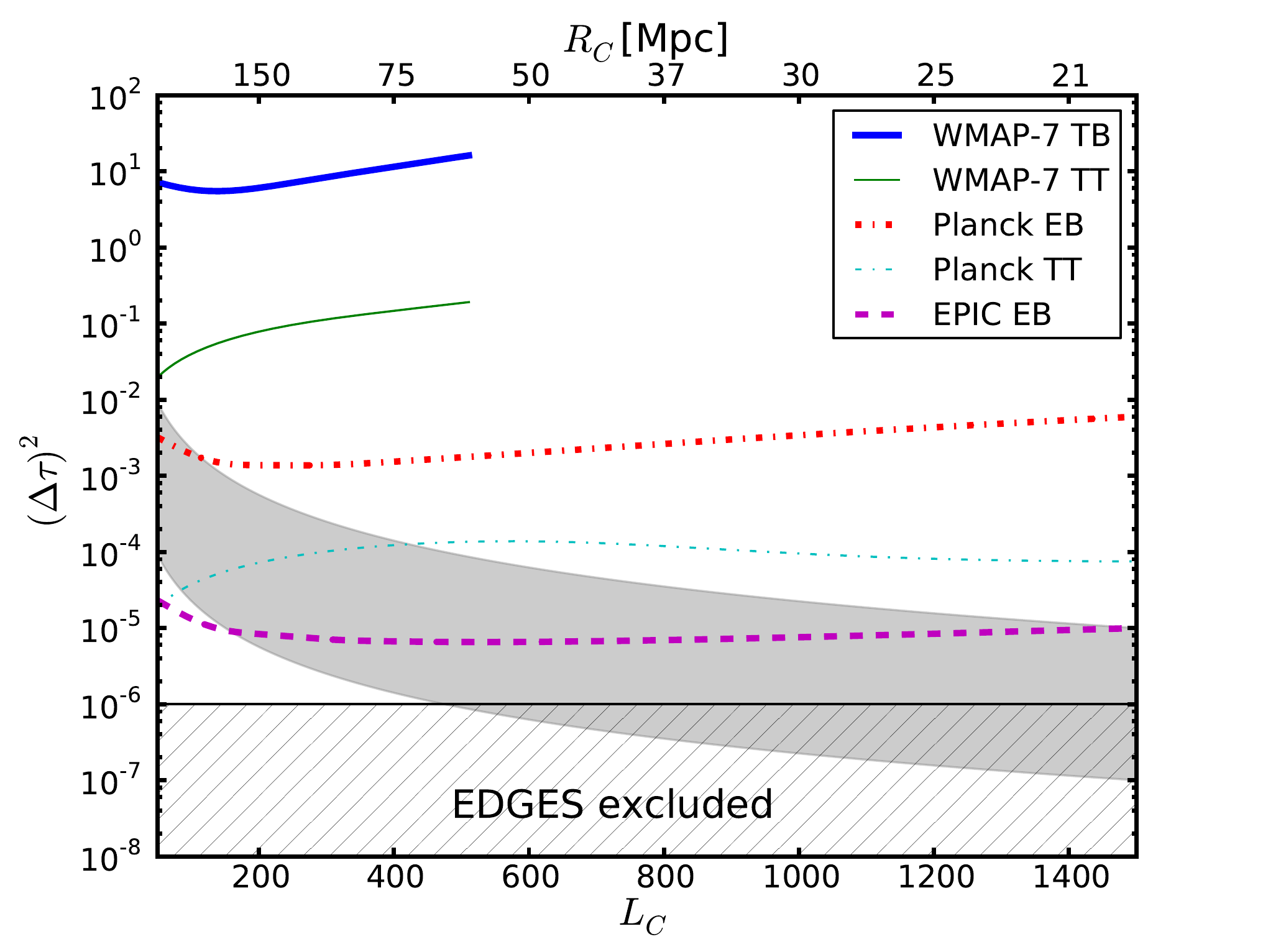}
\caption{Shown are the $1\sigma$ confidence-level upper limits
     from WMAP-7 TT and TB correlations on the amplitude
     $(\Delta\tau)^2$ of the patchy-screening model given in
     Eq.~\eqref{eq:models} as a function of the coherence-scale
     parameter $L_C$. Also shown are sensitivity forecasts for
     experiments with map noise of $27\mu$Karcmin and
     $1\mu$Karcmin, and beam width of $7'$ and $5'$,
     corresponding to Planck and EPIC-like mission, 
     respectively.  The values expected for a reionization
     surface that is crinkled on scales $R_C$ are indicated by
     the gray band.  Also shown is the portion of this parameter space excluded by EDGES
     \cite{Bowman:2012hf}.\label{fig:AvsLc}}
\end{figure}

\textit{Interpretation---}To understand the implications of these
measurements, we consider a simple parametrization of
inhomogeneous reionization in which optical-depth fluctuations
are described by white noise smoothed on angular scales
$\theta_C \equiv \pi/L_C$, or a power spectrum
\beq
     C_L^{\tau\tau} = (4\pi/L_C^2)(\Delta\tau)^2e^{-L^2/L_C^2},
\label{eq:models}
\eeq
shown in Fig.~\ref{fig:models} for several values of $L_C$.
We constrain the parameters $\Delta\tau$ and $L_C$ using
the minimum-variance estimate \cite{Gluscevic:2010vv} for the amplitude,
\beq
     \widehat{(\Delta\tau)^2} = (\sigma[(\Delta\tau)^2])^2
     \sum_L C_L^{\tau\tau\text{,fiducial}}\widehat
     C_L^{\tau\tau}/\text{var}(\widehat C_L^{\tau\tau}),
\label{eq:Afit}
\eeq
where 
\beq
   \left(\sigma[(\Delta\tau)^2]\right)^{-2} = \sum_L
   (C_L^{\tau\tau\text{,fiducial}})^2 / \text{var}
   (\widehat C_L^{\tau\tau})
\label{eq:sigmaA}
\eeq
is roughly the inverse-variance with which $(\Delta\tau)^2$ can be measured, $\text{var}(\widehat C_L^{\tau\tau})$ is the variance of the
power spectrum, estimated from a suite of simulations with no
patchy screening, and $\widehat C_L^{\tau\tau}$ are the unbinned
measurements from WMAP-7 maps. Since the results
are consistent with no signal, the variance provides a
constraint on $(\Delta\tau)^2$, which we show as a function of
the model parameter $L_C$ in Fig.~\ref{fig:AvsLc}. We also show in Fig.~\ref{fig:AvsLc} the upper limit to $(\Delta\tau)^2$ inferred from upper limits to the power of  TT modulation \cite{Hanson:2009gu}.
Given that the mean optical depth is known to be $\tau \sim
0.1$, it is clear that our bounds $\Delta\tau \lesssim 1$, from
$TB$ is far from constraining, and that $\Delta\tau \lesssim
0.1$, from $TT$ is at best marginally constraining.

Eq.~(\ref{eq:models}) describes what happens if every point in
the Universe goes suddenly from neutral to ionized, but with a
reionization surface that is crinkled on a comoving scale of
$R_C \simeq 200\,\text{Mpc} \, (L_C/150)^{-1}$.  This smoothing
scale, or bubble size, corresponds at a reionization redshift
$z_r\sim 10$ to a redshift interval $\Delta z \sim R_C z_r^{1/2}
\Omega_m^{1/2} H_0/c$.  Since the optical depth scales with the
reionization redshift as $\tau\propto z^{3/2}$, we find that a
bubble size $R_C$ induces an optical-depth fluctuation $\Delta
\tau \simeq 0.01(R_C/200\,\text{Mpc})$.  There is thus a rough
scaling, $(\Delta\tau) \sim 0.01 \,(L_C/150)^{-1}$, between the
optical-depth-fluctuation amplitude and the correlation
multipole $L_C$ for the crinkly-surface model, represented by a thick band (to indicate roughly the theory uncertainty) in
Fig.~\ref{fig:AvsLc}.  Fig.~\ref{fig:AvsLc} also shows the
expectations \cite{Dvorkin:2008tf,Gluscevic:2009mm} for the
sensitivities of Planck and EPIC
\cite{Baumann:2008aj}.  Also shown is a constraint for this
crinkly-surface model from the EDGES constraint, $\Delta z
> 0.06$ (at $95\%$ confidence) \cite{Bowman:2012hf}, from the
all-sky redshifted 21-cm spectrum.

A wider range of reionization scenarios can be described by
a ``Swiss cheese'' model in which bubbles of size
$R_C$ are spread over a larger redshift range
\cite{Gruzinov:1998un}, so that each line of sight crosses, on
average, $N$ bubbles.   The rms optical-depth fluctuation in
Eq.~(\ref{eq:models}) would, for fixed $R_C$, then be reduced by
a factor $N^{1/2}$, relative to the crinkly-surface model.  Thus,
both the gray shaded area, and the ``EDGES excluded'' regions in
Fig.~\ref{fig:AvsLc} would be reduced by $N^{1/2}$. Note that kSZ fluctuations should increase in sensitivity as $N$ increases \cite{Zahn:2011vp,Gruzinov:1998un} to complement the reduced sensitivity of patchy screening in this limit.

\textit{Conclusions---}The directional dependence of the optical
depth $\tau(\hatn)$ encodes information about the morphology of
the ionized regions during the epoch of reionization. Here we
have used WMAP-7 temperature and polarization data to derive a
bound on the individual multipoles of the optical-depth
power spectrum up to $L=512$, or bubble sizes larger
than $\sim 60$ Mpc comoving.  We then interpreted these null results in terms of
a bound on an rms optical-depth fluctuation $\Delta\tau$ in a
model of white-noise fluctuations with coherence angle
$\theta_C$.  While
the bound derived proves to be too weak to constrain realistic
models, and probes bubble sizes larger than those ($R_C\lesssim
10$~Mpc) favored in current reionization models, our result
provides a proof of principle that such analyses can be carried
out with future data.  We then note that data from the
forthcoming Planck satellite and from a subsequent post-Planck
project should approach the realistic parameter
space.  Before such optical-depth-fluctuation searches are
carried out in the future, though, several issues will need to
be understood.  For example, the estimator in
Eq.~\eqref{eqn:taulm_estimator} has the same parity as that for
the lensing potential \cite{Hirata:2003ka,Gluscevic:2009mm}, and
further modeling of the $\Delta\tau$ and lensing signals, and/or
de-lensing of the CMB, will be necessary for a CMB detection in
the optical-depth fluctuation with Planck \cite{Planck} or
future-generation experiments.

\textit{Acknowledgments---}The authors thank Gil Holder for useful discussions. This work was supported by the Simons Foundation at Caltech, DoESC-0008108 and NASA NNX12AE86G at Caltech, and a CITA National Fellowship at McGill. Some of the results in this paper have been derived using HEALPix \cite{Gorski:2004by}.



\begin{thebibliography}{}

\bibitem{Kamionkowski:1993aw} 
  M.~Kamionkowski, D.~N.~Spergel and N.~Sugiyama,
  Astrophys.\ J.\  {\bf 426}, L57 (1994)
  [astro-ph/9401003];
  M.\ Fukugita and T.\ Kawasaki,
  Mon. Not. R.\ Astron.\ Soc.\ {\bf 269}, 563 (1994); 
  P.~R.~Shapiro, M.~L.~Giroux and A.~Babul,
  Astrophys.\ J.\  {\bf 427}, 25 (1994);
  M.~Tegmark, J.~Silk and A.~Blanchard, 
  Astrophys.\ J.\  {\bf 420}, 484 (1994)
  [astro-ph/9307017].

\bibitem{reionquasar}
  R.~H.~Becker {\it et al.}  [SDSS Collaboration],
  Astron.\ J.\  {\bf 122}, 2850 (2001)
  [astro-ph/0108097].
  X.\ Fan {\it et al.}, Astron.\ J. {\bf 132}, 117 (2006).

\bibitem{Oh:2004rm} 
  S.~P.~Oh and S.~R.~Furlanetto,
  Astrophys.\ J.\  {\bf 620}, L9 (2005)
  [astro-ph/0411152].

\bibitem{Komatsu:2010fb}
  E.~Komatsu {\it et al.}  [WMAP Collaboration],
  Astrophys.\ J.\ Suppl.\  {\bf 192}, 18 (2011)
  [arXiv:1001.4538 [astro-ph.CO]].

\bibitem{Larson:2010gs} 
  D.~Larson {\it et al.} [WMAP Collaboration],
  Astrophys.\ J.\ Suppl.\  {\bf 192}, 16 (2011)
  [arXiv:1001.4635 [astro-ph.CO]].

\bibitem{Barkana:2000fd} 
  R.~Barkana and A.~Loeb,
  Phys.\ Rept.\  {\bf 349}, 125 (2001)
  [astro-ph/0010468].

\bibitem{Furlanetto:2006jb} 
  S.~Furlanetto, S.~P.~Oh and F.~Briggs,
  Phys.\ Rept.\  {\bf 433}, 181 (2006)
  [astro-ph/0608032].

\bibitem{Pritchard:2011xb} 
  J.~R.~Pritchard and A.~Loeb,
  Rep.\ Prog.\ Phys.\ {\bf 75}, 086901 (2012)
  [arXiv:1109.6012 [astro-ph.CO]].

\bibitem{Bowman:2012hf} 
  J.~D.~Bowman and A.~E.~E.~Rogers,
  Nature {\bf 468}, 796 (2010)
  [arXiv:1209.1117 [astro-ph.CO]].

\bibitem{Zahn:2011vp} 
  O.~Zahn {\it et al.} [SPT Collaboration],
  arXiv:1111.6386 [astro-ph.CO].

\bibitem{Gruzinov:1998un} 
  A.~Gruzinov and W.~Hu,
  Astrophys.\ J.\  {\bf 508}, 435 (1998)
  [astro-ph/9803188];
  L.~Knox, R.~Scoccimarro and S.~Dodelson,
  Phys.\ Rev.\ Lett.\  {\bf 81}, 2004 (1998)
  [astro-ph/9805012].

\bibitem{Dvorkin:2009ah} 
  C.~Dvorkin, W.~Hu and K.~M.~Smith,
  Phys.\ Rev.\ D {\bf 79}, 107302 (2009)
  [arXiv:0902.4413 [astro-ph.CO]].

\bibitem{Dvorkin:2008tf} 
  C.~Dvorkin and K.~M.~Smith,
  Phys.\ Rev.\ D {\bf 79}, 043003 (2009)
  [arXiv:0812.1566 [astro-ph]].

\bibitem{Hanson:2009gu} 
  D.~Hanson and A.~Lewis,
  Phys.\ Rev.\ D {\bf 80}, 063004 (2009)
  [arXiv:0908.0963 [astro-ph.CO]].

\bibitem{Gluscevic:2009mm}
  V.~Gluscevic, M.~Kamionkowski and A.~Cooray,
  Phys.\ Rev.\  D {\bf 80}, 023510 (2009)
  [arXiv:0905.1687 [astro-ph.CO]].

\bibitem{Gluscevic:2012me} 
  V.~Gluscevic, D.~Hanson, M.~Kamionkowski and C.~M.~Hirata,
  arXiv:1206.5546 [astro-ph.CO].

\bibitem{lambda}
	{\tt http://lambda.gsfc.nasa.gov/}

\bibitem{Gluscevic:2010vv} 
  V.~Gluscevic and M.~Kamionkowski,
  Phys.\ Rev.\ D {\bf 81}, 123529 (2010)
  [arXiv:1002.1308 [astro-ph.CO]].

\bibitem{Planck} {\tt http://www.rssd.esa.int/PLANCK}.

\bibitem{Baumann:2008aj} 
  D.~Baumann {\it et al.}  [CMBPol Study Team Collaboration],
  AIP Conf.\ Proc.\  {\bf 1141}, 3 (2009)
  [arXiv:0811.3911 [astro-ph]].

\bibitem{Hirata:2003ka} 
  C.~M.~Hirata and U.~Seljak,
  Phys.\ Rev.\ D {\bf 68}, 083002 (2003)
  [astro-ph/0306354].

\bibitem{Gorski:2004by} 
  K.~M.~Gorski {\it et al.},
  Astrophys.\ J.\  {\bf 622}, 759 (2005)
  [astro-ph/0409513];
  {\tt http://healpix.jpl.nasa.gov}.


\end{thebibliography}
\end{document}